\documentclass[aps,showkeys,showpacs,nosuperscriptaddress,twocolumn]{revtex4}

\usepackage{graphicx}
\usepackage{amsmath}
\usepackage{hyperref}
\usepackage{color}
\usepackage{subfigure}
 
\begin{document}

\title{Combinatorial aspect of fashion}

\author{M.~J.~Krawczyk}
\email{gos@fatcat.ftj.agh.edu.pl}
\affiliation{
Faculty of Physics and Applied Computer Science, AGH University of Science and Technology, al. Mickiewicza 30, PL-30059 Krak\'ow, Poland
}

\author{K.~Ku{\l}akowski}
\email{kulakowski@fis.agh.edu.pl}
\affiliation{
Faculty of Physics and Applied Computer Science, AGH University of Science and Technology, al. Mickiewicza 30, PL-30059 Krak\'ow, Poland
}
 
\date{\today}
 
\begin{abstract}
Simulations are performed according to the Axelrod model of culture dissemination, with modified mechanism of repulsion. Previously,
repulsion was considered by Radillo-Diaz et al (Phys. Rev. E 80 (2009) 066107) as dependent on a predefined threshold. Here the probabilities
of attraction and repulsion are calculated from the number of cells in the same states. We also investigate the influence of some homogeneity,
introduced to the initial state. As the result of the probabilistic definition of repulsion, the ordered state vanishes. A small cluster
of a few percent of population is retained only if in the initial state a set of agents is prepared in the same state. We conclude that the 
modelled imitation is successful only with respect to agents, and not only their features.
\end{abstract}
 
 

\maketitle

\section{Introduction}

Mathematical metaphors and illustrations of social phenomena are used in sociology for tens of years 
\cite{coleman,granovetter,axelrod,schelling}. Still, an interest in these seminal papers 
\cite{granovetter,axelrod,schelling,axel2} has grown considerably only in last ten years \cite{webofscience}. 
This phenomenon can be interpreted in different ways. Certainly, mathematical expressions can be impressive,
as they seem to offer some a priori formalization and logical thinking; this pretence creates some kind of fashion 
what was rightly criticized for a long time \cite{andreski,sokal}. On the other hand, it is possible that 
quantitative considerations will give input to our - necessarily permanent - reconstructions of sociological 
ideas. This belief cannot be verified within sciences themselves (\cite{simmel}, p.19); yet it opened 
the way for massif efforts of scientists to contribute into mathematical and computational sociology.
Below we continue this trend along the direction pointed by Robert Axelrod in his model 
of culture dissemination \cite{axel2}. \\

In accordance with the mathematical trend mentioned above, "culture" in this model is described as a set of 
symbols, and human beings (agents) as chains of length $F$. In each of chain cells a symbol is written, one 
out of $q$ possible values. $F$ and $q$ are model parameters. Agents are placed in a square lattice, and interact 
only with their four nearest neighbours (von Neumann neighbourhood). Probability of an interaction between 
two agents is proportional to the number $k$ of chain cells, where both of them have the same symbols. To give an
example, suppose that $F$=5, $q$=7 and the chains of two neighbours are: 13245, 23575. In this example, 
the symbols in second and fifth cells of one chain are in the same state that in the other chain; hence the 
interaction probability $p=k/F$ is 2/5. As the consequence of the interaction, the number $k$ of cells in the same state 
increases by one \cite{axel2}. The rationale of the model is as follows. Symbols in chain cells represent cultural features.
Agents who do not share any feature are supposed to ignore each other. However, if at least one feature 
is the same for two neighbouring agents, their contact becomes possible and this leads to a further unification. 
The so-called active bonds between neighbours with $0<p<1$ is the fuel of changes, where nodes always attract each 
other in the state space. In the stationary absorbing state, nearest neighbours are either identical or completely different.\\

To explain our motivation, we refer to two papers along the same direction. In 2000, a phase transition has been 
identified in the model \cite{castellano}. Below some value of $q$, say $q*$, the system tends to a homogeneous 
phase, where all agents are described by the same chains of symbols. For $q>q*$, this ordering vanishes. In 2009,
a kind of threshold-dependent repulsion has been introduced to the model \cite{radillo}. Namely, the number $k$ of 
identical cells in a pair is compared with some threshold $\gamma$. Once $k>\gamma F$, the interaction is attractive. 
However, if $k<\gamma F$, the interaction leads to a decrease of $k$ by one; this is equivalent to a repulsion in 
the state space. As a consequence, the value of $q*$ is strongly reduced. Also, a small (about 2 percent) cluster 
of agents in the same state is observed near some $q^+$ well above $q*$. \\

The very idea of repulsion was used recently \cite{huet} to generalize the bounded confidence model of public opinion \cite{weisbuch}.
In this model, agents attract in the continuous space of issues (say, at the plane of safety and welfare), if their
initial positions are closer to each other than some threshold. The repulsion means that once their distance along 
the axis of a leading issue is larger than another threshold value, their coordinates along the other axis get 
different as well \cite{huet}. Our position is that in the multidimensional space of symbols \cite{castellano}, the 
idea of threshold ceases significance. Two agents, once they met, have no time to investigate the states of all their 
cells. Instead, they sample one cell, i.e. one feature, and decide on the basis of acquired information. In other words, 
they attract with the probability $p=k/F$ and they repulse with the probability $1-p$. Another new element of our work 
is that we take into account the initial state. Namely, some percentage $d$ of chain cells is set to be endowed with 
the same symbols. This is done for each positions, $1,\hdots,F$. The rate $d$ and the method of this preparation appears to 
be relevant for the outcome. \\

The next section provides algorithmic details of our approach, including four different methods of preparation of initial 
states. Section 3 is devoted to our numerical results, an almost complete destruction of the homogeneous state being the 
most important. Final conclusions - with an attempt to interpret the results within the phenomenon of fashion - are given 
in the last section. \\

\section{Calculations}

Agents are placed at nodes of a square lattice $L\times L$, with periodic boundary conditions.  Each agent is endowed with 
a chain of $F$ cells, with a symbol placed at each cell. Symbols can take one of $q$ values. Initially, the values of all symbols 
can be set randomly, with uniform probability. Alternatively, the initial state can be prepared as follows. For each cell $i=1,\hdots,F$, 
we find the symbol value $X_i$ which appears in this cell most frequently. Then we select randomly $dL^2$ agents and we overwrite 
$X_i$ at their $i$-th cells. Here we apply four methods:\\
A1. For each cell $i$, agents where the cell is overwritten are selected separately and randomly.\\ 
A2. For each cell $i$, a lattice node is selected separately. Agents where the cell is overwritten are selected as close to this node as possible.\\
B1. A set of agents is selected  randomly, and their all cells are overwritten.\\
B2. A node is selected randomly. All overwritten cells belong to agents as close to the selected node as possible. \\

The simulation is performed as follows. At each time step, a pair 
of neighbouring agents is selected randomly. One by one, the values of their symbols in the respective cells are compared. 
The probability $p$ of attractive interaction is set to $k/F$, where $k$ is the number of cases where the same symbols 
are found for both agents in the same cell. In the case of attraction, a cell is found where symbols for the agents are 
different, and the value from one agents is copied for another agent. In this way, $k$ for this pair of agents is increased by one. 
If the interaction is not attractive, it is repulsive. This means that $k$ is reduced by one with probability $1-k/F$. Namely,
a cell is found where symbols for the agents are the same, and the symbol for one agent is changed. \\

\section{Results}

The results presented below are obtained for $L=50$ and $F=10$. They are confirmed by less complete calculations for $L=32$, $F=10$.
For a comparison, a set of calculations are performed also for the case without repulsion. In this case the interaction is either attractive
(with probability $k/F$) or has no consequences. We start with the results without repulsion. They do depend on $d$ and on the method 
of preparation of the initial state. These results are shown in Fig. 1 a-d. As we see, for $d=0$ the ordered (homogeneous) state appears
below $q=50$. This agrees with the result in Fig. 2 of \cite{klemm} without noise. Further, when A1 or A2 is applied, the same ordered 
state appears for $q<50$ for all values of $d$; for larger $q$, $d>0.1$ (A1) or $d>0.25$ (A2) is large enough to ensure full ordering. 
In other words, prepared order does not reduce order. This natural result is not necessarily true for B1 and B2. There, ordering is 
weakened already for $q>30$ (B1) or $q>40$ (B2) if $d>0.4$. \\

\begin{figure}[htbp!]
\centering
\subfigure[A1]{
\includegraphics[scale=.65]{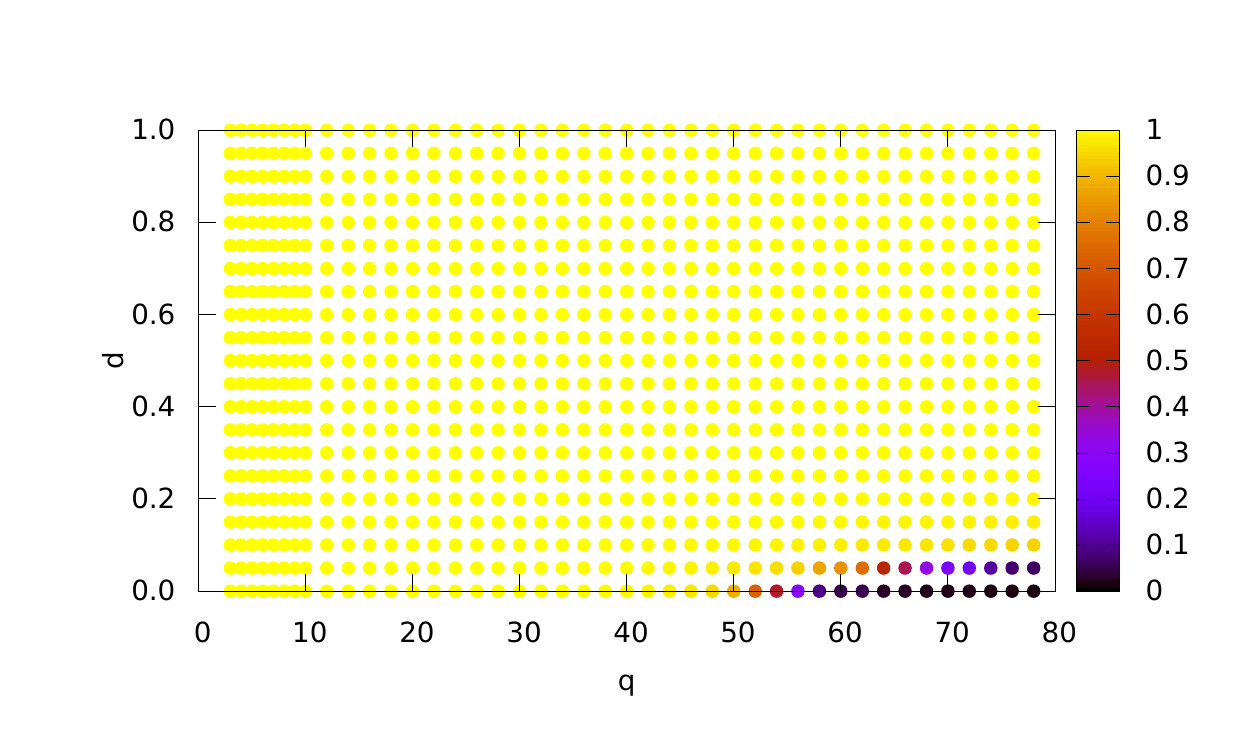}
\label{fig:1A}
}
\subfigure[A2]{
\includegraphics[scale=.65]{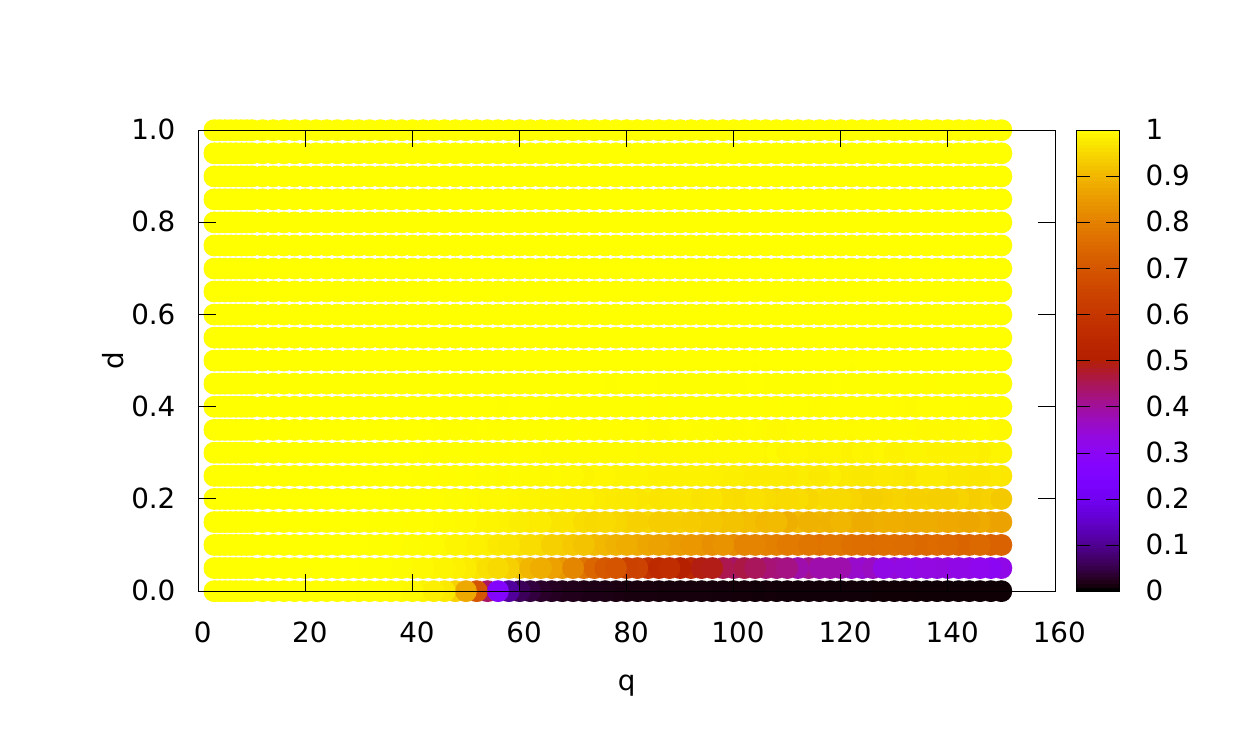}
\label{fig:1B}
}\\
\subfigure[B1]{
\includegraphics[scale=.65]{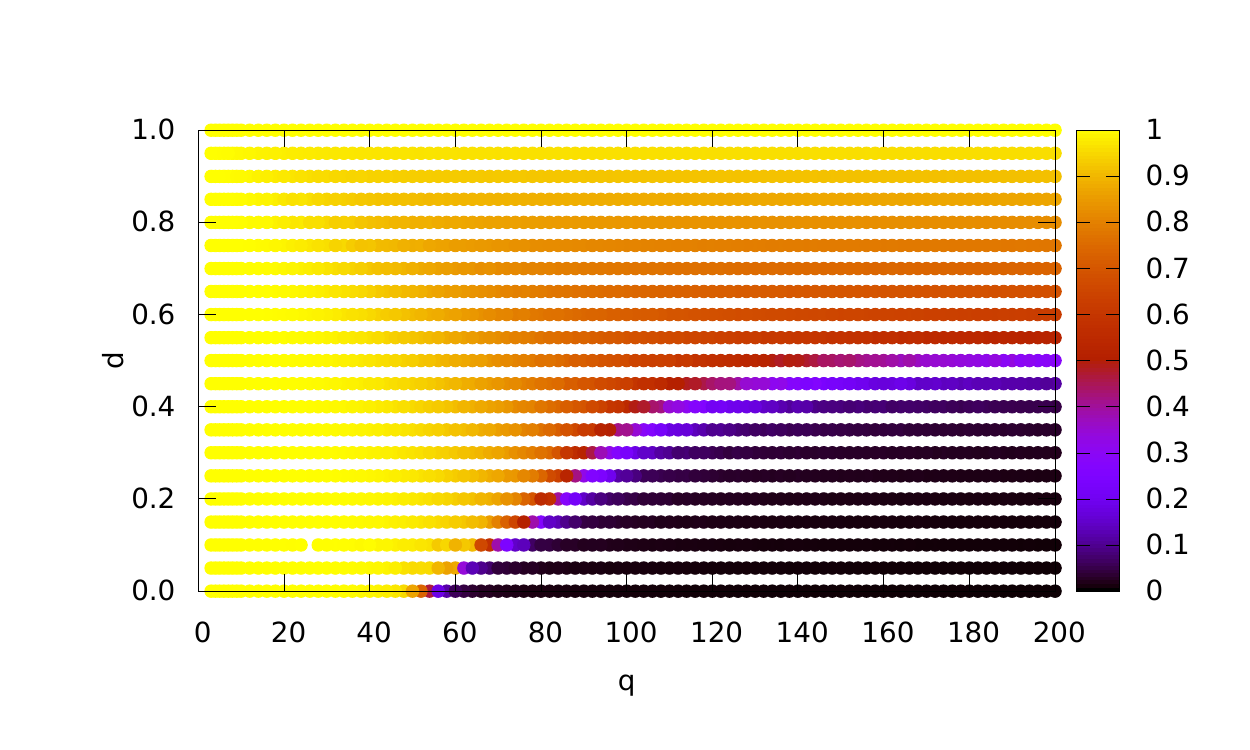}
\label{fig:1C}
}
\subfigure[B2]{
\includegraphics[scale=.65]{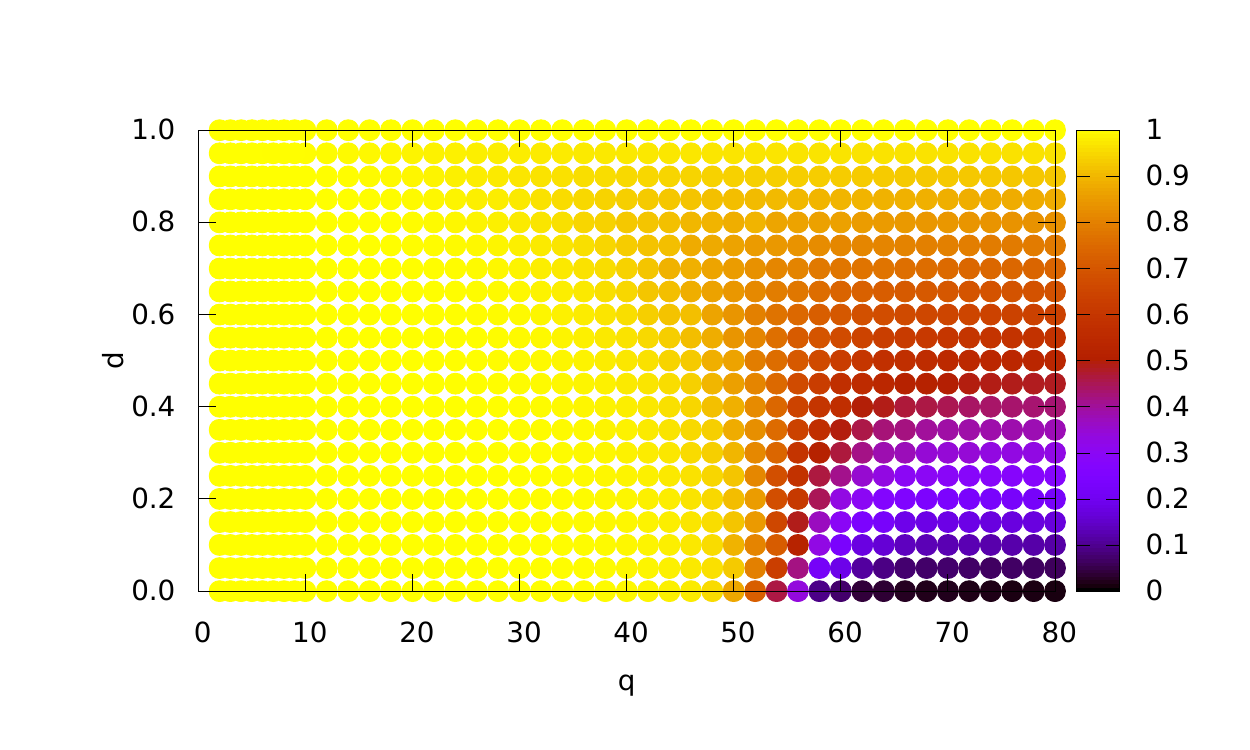}
\label{fig:1D}
}
\label{norep}
\caption[]{The size $max$ of the maximal cluster against the percentage $d$ of modified nodes for different values of the parameter $q$, for 
the four options (a) A1, (b) A2, (c) B1 and (d) B2 of preparation of initial state. The calculations are made for the case without repulsion.}
\end{figure}
 
When the repulsion is turned on, the results are completely different. First of all, the ordered phase disappears for all values of $q$ and $d$, 
except the trivial case $d=1$ (not shown). Both when A1 and A2 are applied, the size of the maximal cluster does not exceed 0.002 (five nodes) even for d=0.95. 
Basically, the ordered phase never appears also for the procedures B1 and B2. However, for B1 we observe a maximal cluster with a maximum 
near $d=0.5$, which increases from about 0.005 (10 nodes) for $q=10$ to about 0.02 (50 nodes) for $q=80$. With B2, a similar but more fuzzy 
maximum of the maximal cluster of the same order is found for much larger $q$ (near 200). The results are shown in Fig. 2.\\

\begin{figure}[htbp!]
\centering
\subfigure[A1]{
\includegraphics[scale=.65]{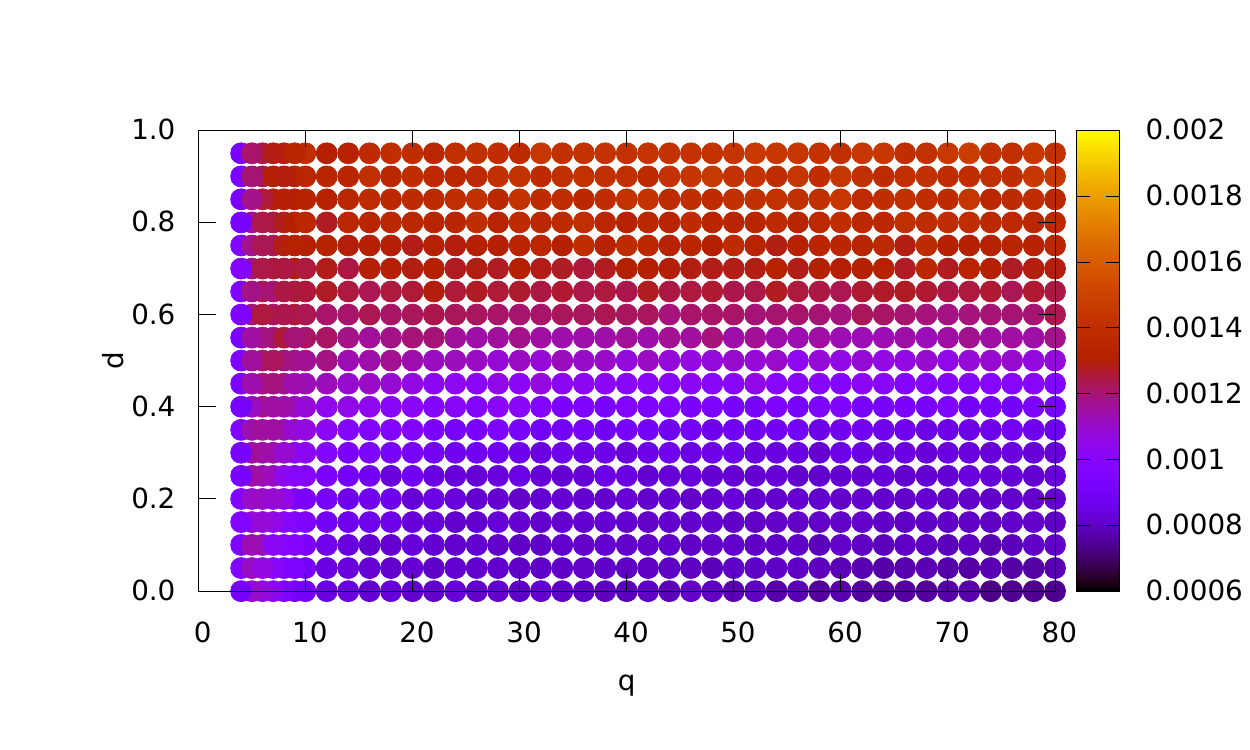}
\label{fig:2A}
}
\subfigure[A2]{
\includegraphics[scale=.65]{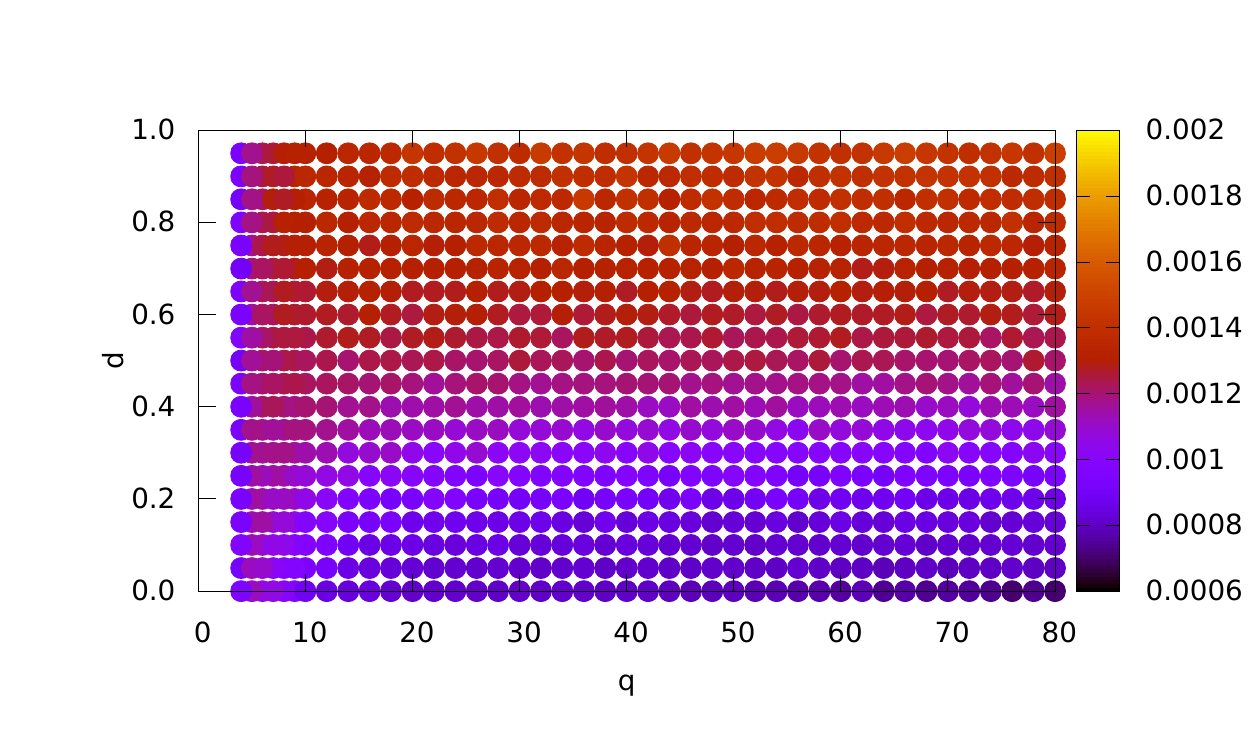}
\label{fig:2B}
}\\
\subfigure[B1]{
\includegraphics[scale=.65]{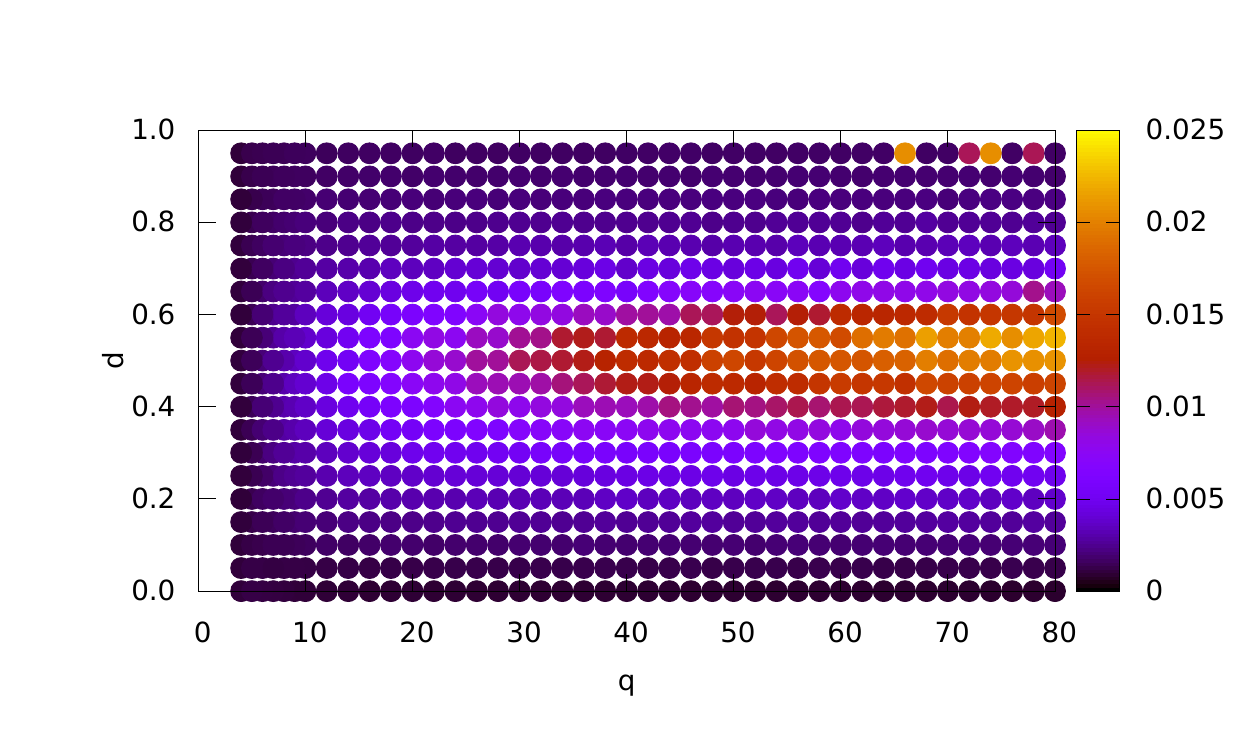}
\label{fig:2C}
}
\subfigure[B2]{
\includegraphics[scale=.65]{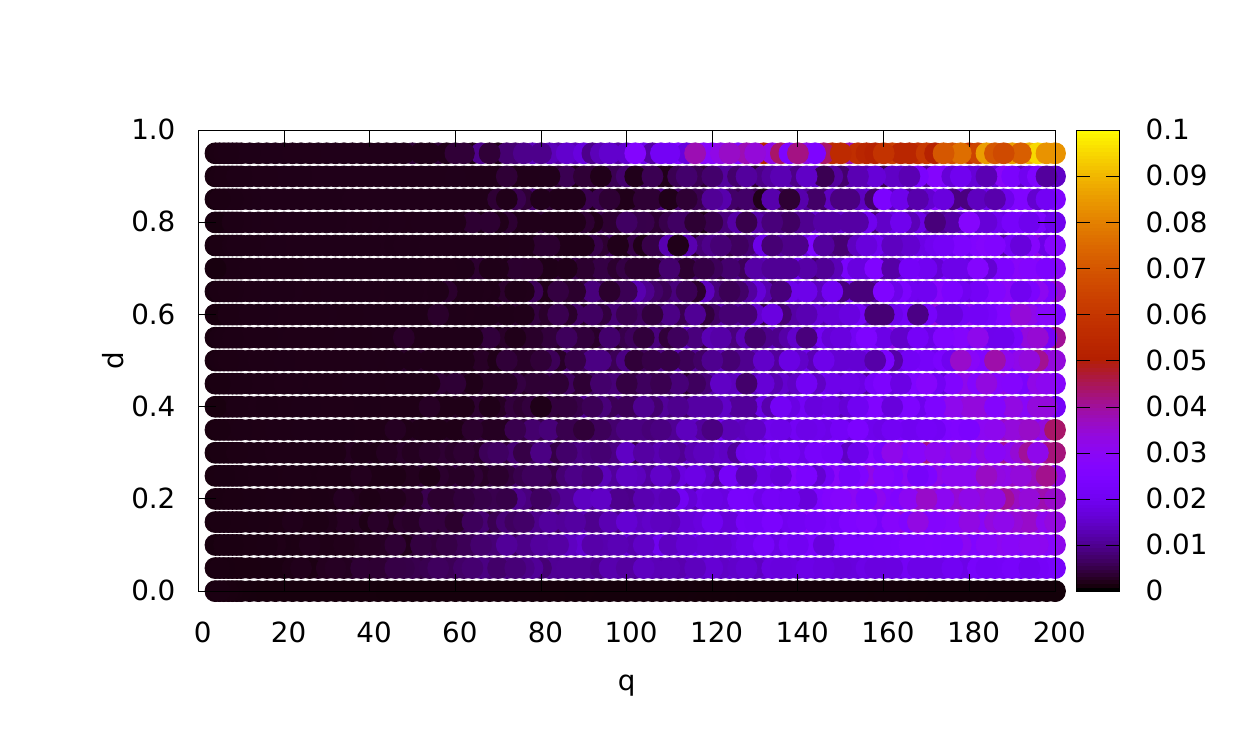}
\label{fig:2D}
}
\label{fig:rep}
\caption[]{The size $max$ of the maximal cluster against the percentage $d$ of modified nodes for different values of the parameter $q$, for 
the four options (a) A1, (b) A2, (c) B1 and (d) B2 of preparation of initial state. The calculations are made for the case with repulsion.}
\end{figure}
 
To get some insight into these results, we inspected also the time dependence of the number of active bonds. In general, two kinds of curves are 
obtained, depending on if the initial state is near or far to the final absorbing state. As a rule, the latter option is characterized by 
a clear increase of the number of active bonds after some transient time. However, sometimes it happens that - seemingly by chance - the absorbing
state is found before the above mentioned increase. Once the number of active bonds reaches zero, the system cannot evolve anymore. Such a
metastable absorbing state may occur when repulsive interaction is present or not, and when the final state is ordered state or not. Examples 
are shown in Fig. 3.\\

\begin{figure}[htbp!]
\centering
\subfigure[No repulsion, B2]{
\includegraphics[scale=.65]{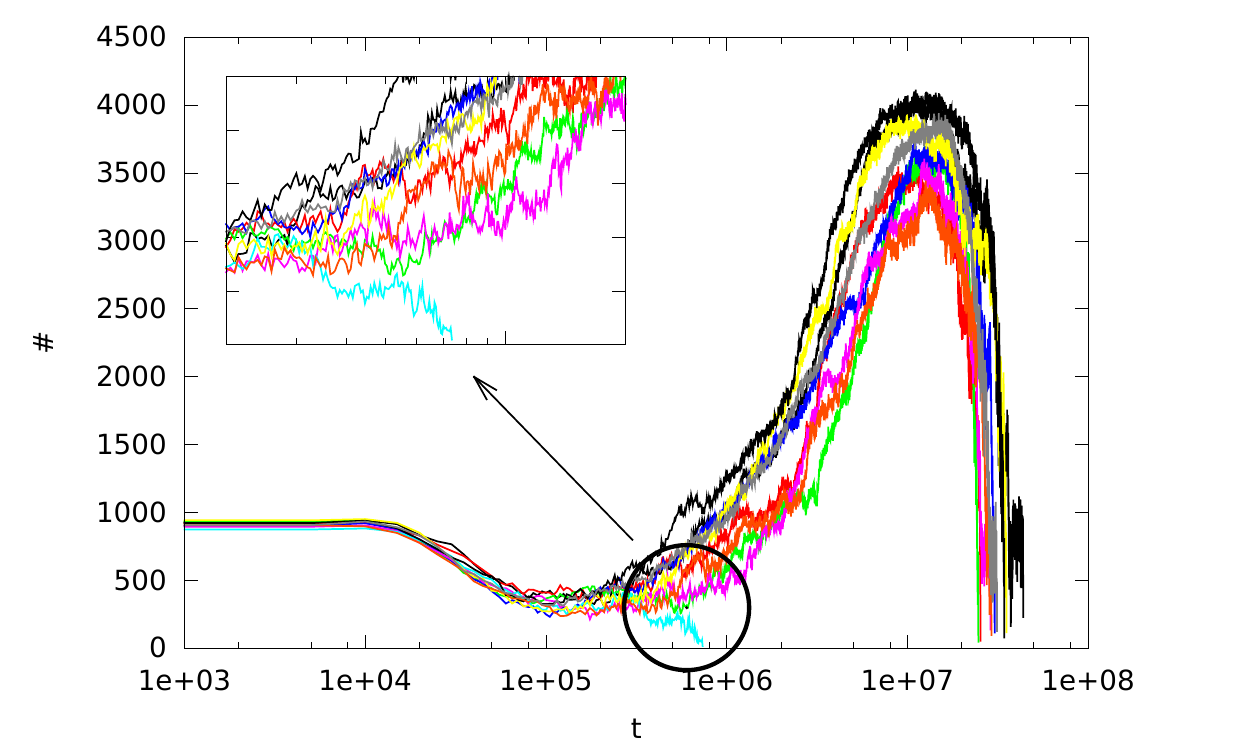}
\label{fig:subfig3A}
}
\subfigure[Repulsion, B2]{
\includegraphics[scale=.65]{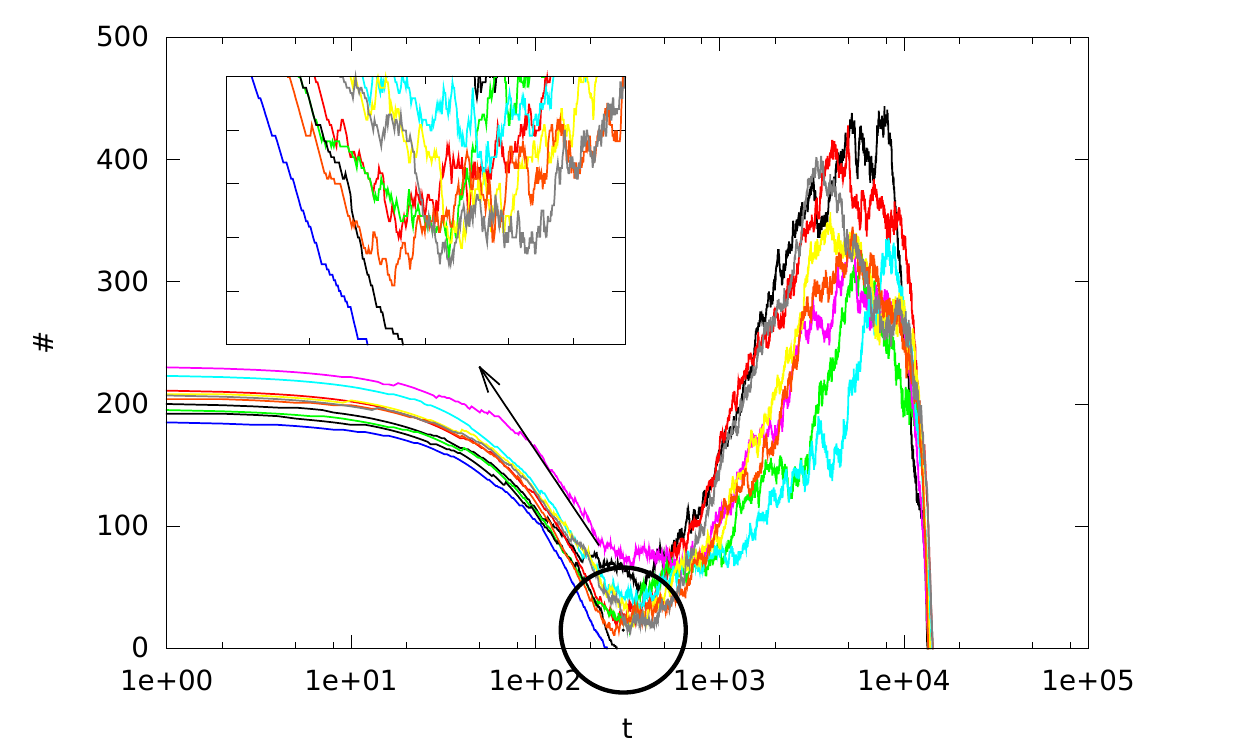}
\label{fig:subfig3B}
}\\
\label{fig:time}
\caption[]{Time dependence of the number $\#$ of active bonds for the case (a) without repulsion, $d=0$, option B2, and (b) with repulsion, $d=0.2$, option B2. In both cases,
most trajectories show a maximum due to the change of the size $max$ of the largest cluster. However, in both cases it is possible that the trajectory
happens to be blocked in an absorbing state where $max$ is close to its initial value. In the pictures above, (a) one trajectory ends at $d=0.1$ while 
the remaining nine trajectories end with $d>0.9$, (b) two trajectories end at $d=0.2$ while the remaining eight trajectories end at $d<0.002$. These 
exceptional trajectories are shown in the insets.}
\end{figure}

\section{Discussion}

The most important result is that in its proposed form, the repulsion destroys the ordered phase. This result can be interpreted in two ways,
as negative ("divided we fall" \cite{sznajd}) or positive ("diversity is preserved" \cite{macy}). In any case, our results indicate that the 
details of the repulsion mechanism do matter. This is seen when we compare our results with those of \cite{radillo}, where the decision to attract 
or to repulse was dependent on a prescribed threshold $\gamma$. Namely, if $k/F<\gamma$, the interaction was repulsive. This kind of interaction 
is less stochastic in the sense that the decision does not rely on any probability. We note that in \cite{radillo}, the ordered phase was 
preserved for small $q$. In our model of repulsion, this ordered phase does not appear.\\

There is still some correspondence between the results for these two mechanisms of repulsion. Namely, a small ordered cluster was found in \cite{radillo}
which appears above the threshold $q*$. Its size reported was the largest for the threshold $\gamma=0.2$, but it was not more than $0.02L^2$. 
This effect is similar to ours, as shown in Fig. 2c, with an important difference that the maximum found by us does not decrease with $q$ and appears only for partial 
initial ordering. It seems that the origin of the maximum is combinatorial in both models. Driven by the subject of the Axelrod model, we are tempted 
to search its analogy in cultural world. According to the seminal paper \cite{axelrod}, one of possible applications of the symbols is fashion. 
Voting for this option, we are willing to interpret cells as details of dress, as tie, glasses, shoes or handbag. This interpretation makes large 
values of $q$ (even one hundred!) understandable. Also, we can agree that the largest cluster of the order of two percent of the interacting 
population is realistic. On the contrary, the interpretation of languages, mentioned in \cite{castellano}, seems not appropriate, as nowhere 
in the world we have fifty languages as equivalent options.\\

As we remarked in the Introduction, it seems to us that the version of repulsion proposed by us is more close to a sociological reality, than 
a mechanism which depends on a pre-defined threshold \cite{radillo}. Our second modification - a generalization of the model - is to prepare an initial 
state. We believe that a perfectly disordered society does not exist, and if it does, the nature of this disorder is not interesting for social sciences.
Driven by this belief, we made the generalization in four different ways. The results are puzzling in the sense that the most interesting effect -
the small cluster - is present for options B1 and B2, and neither for A1, nor for A2. This means that the spatial distribution of the initially modified 
nodes is not relevant. This is odd, when we remember that the interaction range is limited to the nearest neighbours. What is relevant is the 
concentration of all modified cells in the same agents. This result, when translated to sociological reality, reads that fashion is a result of 
imitation agents, and not their particular feature. The effect is not difficult to be recognized in real world. What is puzzling is that it has 
also a combinatorial aspect. Indeed, it is only this aspect of culture which can be captured in the Axelrod model.\\

\begin{acknowledgments}
The results have been presented during the 6-th Symposium "Physics in Economy and Social Sciences" in Gda\'nsk, April 19-21, 2012.
One of the authors (K.K.) is grateful to the Organizers for their kind hospitality. The calculations were made at the ACK Cyfronet Center. 
The research is partially supported within the FP7 project SOCIONICAL, No. 231288.
\end{acknowledgments}

 \end{document}